\documentclass[letterpaper,10pt,times,mathptm,psfig,conference]{IEEEtran}
\IEEEoverridecommandlockouts
\usepackage{cite}
\usepackage{amsmath,amssymb,amsfonts}
\usepackage{xspace}       % needed by \name macros (was pulled in via algorithm2e)
\usepackage[normalem]{ulem}
\usepackage{amssymb}
\usepackage{amsmath}
\usepackage{graphicx}
\usepackage[caption=false,font=footnotesize]{subfig}
\usepackage{textcomp}
\usepackage{xcolor}
\usepackage{hyperref}
\usepackage{epsfig, amsmath, amsthm, amssymb, wrapfig, bm} % subfigure removed to see if it fixes an issue
\usepackage{enumitem}

\usepackage{pifont}
\usepackage{threeparttable}
\usepackage{multirow}
\usepackage{tabularx}
\usepackage{algorithm}
\usepackage[noend]{algpseudocode}

\DeclareMathOperator{\Com}{Com}

\newcommand{\limit}{\mathrm{limit}}

\def\BibTeX{{\rm B\kern-.05em{\sc i\kern-.025em b}\kern-.08em
    T\kern-.1667em\lower.7ex\hbox{E}\kern-.125emX}}

\newcommand{\names}{{\textsc{\scriptsize{GridShield}}}\xspace}
\newcommand{\name}{{\textsc{\small{GridShield}}}\xspace}
\newcommand{\namet}{{\textsc{GridShield}}\xspace}

\newcommand{\khalil}[1]{}

\newcommand{\kafi}[1]{\textcolor{black}{#1}}

\author{
\IEEEauthorblockN{
Abdullahil Kafi\IEEEauthorrefmark{1},
Md Zahidul Islam\IEEEauthorrefmark{2},
and Alvi Ataur Khalil\IEEEauthorrefmark{1}
}
\IEEEauthorblockA{
\IEEEauthorrefmark{1}\textit{Computer Science, School of Computing},
\IEEEauthorrefmark{2}\textit{School of Electrical, Computer, and Biomedical Engineering}\\
Southern Illinois University Carbondale, Carbondale, IL, USA\\
Email: \{abdullahil.kafi, mdzahidul.islam, a.khalil\}@siu.edu
}
}

\begin{document}
\title{\namet: Verifiable Grid Feasibility and Manipulation Resistance in Privacy-Preserving Peer-to-Peer Energy Trading}

\maketitle
\thispagestyle{empty}
\pagestyle{empty}

\begin{abstract}
Blockchain-based peer-to-peer (P2P) energy markets let neighbours trade directly; however, concentrate unchecked power in the clearing operator. Most prior work addresses only part of this: network-aware clearing respects grid constraints yet operates on plaintext allocations, while privacy-preserving designs hide bids yet either drop the network constraint, enforce it only inside an unauditable computation, or recover it at the cost of a per-circuit trusted setup. Neither considers what happens if an adversary captures the operator's clearing privilege to reorder trades for profit or curtail a rival. We present \name, a P2P energy-trading system that clears sealed bids with an order-independent batch program, removing the welfare a compromised operator can extract by reordering trades, and proves the result grid-feasible via an aggregated congestion-feasibility proof (ACFP). Exploiting the additive homomorphism of Pedersen commitments, ACFP reconstructs each grid constraint directly from per-bus commitments, reducing feeder-wide feasibility verification over any public grid constraints to a single aggregated range proof over the hidden allocation, with a reveal-and-recompute challenge that catches feasible yet unfair clearings. We introduce congestion-responsibility fairness (CRF), a metric scoring curtailment against each party's physical responsibility for congestion rather than consumption. We evaluate \name against three prior P2P energy-trading designs across four feeders, including a 240-node Midwest feeder with real hourly loads, on a live Hyperledger Fabric network benchmarked with Hyperledger Caliper. \name proves an entire round feasible with a few-kilobyte proof and no trusted setup, rejects infeasible allocations two of the three compared designs accept, and sustains $\approx$ 1064 tx/s.

\end{abstract}


%%%%%%%%%%%%%%%%% Introduction %%%%%%%%%%%%%
\vspace{-6pt}
\section{Introduction}
\label{sec:introduction}

\vspace{-4pt}

Distributed energy resources, rooftop photovoltaics, home batteries, and controllable loads, are turning consumers into prosumers who both produce and consume power. Peer-to-peer (P2P) local energy markets let them trade surplus directly over short clearing intervals, improving the utilization of local generation and deferring costly grid reinforcement~\cite{sousa2019peer}. Such markets are increasingly settled on distributed ledgers as a trustless setup~\cite{andoni2019blockchain}.
Unlike a financial marketplace, however, an energy market is bound by physics: power flows over shared distribution feeders under hard physical limits, thermal ratings and voltage bounds among them, therefore a valid market outcome must also be network-feasible~\cite{guerrero2018decentralized}.

That combination creates an authority the market-design literature has largely overlooked. In a permissioned consortium, the distribution system operator (DSO) typically both operates the ordering service (a service that orders transactions into a sequence before they are added to the blockchain) and clears the market, and when a feeder is congested: the scarce resource is room on the wire. 
The DSO therefore decides who is prioritized first, who is curtailed from trading, and who pays the resulting congestion charge. \kafi{This concentration of privilege is itself an attack surface. If the operator is compromised or coerced, it can manipulate welfare, much like the maximal extractable value (MEV) that ordering power yields in decentralized finance~\cite{daian2020flash}.} Privacy compounds the problem: once bids are hidden to protect prosumers, a curtailed participant cannot distinguish a fair loss from covert favoritism.

\vspace{-2pt}

Transaction-ordering abuse was formalized as MEV by Daian~\textit{et al.}~\cite{daian2020flash}, who demonstrated front-running, back-running, and reordering attacks in decentralized exchanges. The standard countermeasures, fair-ordering consensus~\cite{kelkar2023themis} and frequent batch auctions~\cite{budish2015high}, remove the advantage of speed and position, however remain anchored to financial payoffs; the contested resource is always blockspace, never physical infrastructure. On the energy side, decentralized market-clearing formulations~\cite{sorin2018consensus,guerrero2019trading} embed a linearized power-flow model into the matching logic so that admitted trades respect line limits, with distribution locational marginal pricing the standard tool for pricing the resulting congestion~\cite{bai2017distribution}; permissioned blockchains such as Hyperledger Fabric~\cite{androulaki2018hyperledger}, whose ordering service concentrates sequencing power in a known operator, provide the settlement substrate. Because bids and consumption traces are privacy sensitive, a parallel literature hardens these markets cryptographically: zero-knowledge proofs of bid validity~\cite{hou2022privacy}, secure computation over secret-shared bids~\cite{mitrea2023privacy,zhou2023bidirectional}, or succinct proofs of a feasibility circuit~\cite{kim2026zero}.

Each line, however, concedes what the other provides. Network-aware clearing enforces line limits on a \emph{plaintext} allocation~\cite{sorin2018consensus} whose compliance the operator asserts, so a participant who cannot see the dispatch has no way to check it. Privacy-preserving designs hide the bids; however, either drop the network constraint altogether~\cite{abidin2016mpc}, enforce it only inside a computation no non-participant can audit, or recover public verifiability at the cost of a per-circuit trusted setup. \kafi{Neither line defends against exploitation of the clearing privilege itself, so a compromised operator that reorders trades for profit or curtails a rival goes undetected.} Running two of these designs on our data, we find their verifiers accept allocations loading a conductor to more than its thermal rating. The closest work on the attacker side studies \emph{participant-side} economic gaming, strategic bidding~\cite{jahns2023prevention} and inc-dec gaming~\cite{ehrhart2026congestion}, rather than \emph{infrastructure-side} ordering by the operator, and MEV research never treats physical grid capacity as the contested resource. This paper occupies that empty intersection.

\vspace{-2pt}

We propose \name, a P2P energy-trading system that closes this gap by clearing sealed bids through an order-independent batch program, making the outcome a function of the bid \emph{set} and eliminating gains from transaction reordering. It then proves that the resulting allocation satisfies the grid constraints without revealing the allocation itself. That proof, an aggregated congestion-feasibility proof (ACFP), rests on a single structural observation: every grid constraint is a linear function of the nodal injections, and Pedersen commitments~\cite{pedersen1991non} are additively homomorphic, so a commitment to each constraint can be rebuilt from the per-bus commitments and the public constraint matrix, and feeder-wide feasibility, over \emph{any} set of linear constraints, collapses to one aggregated range proof with no arithmetic circuit and no trusted setup. Our contributions are three-fold:

\vspace{-2pt}
\begin{itemize}[leftmargin=*, labelindent=0pt]
\item We propose an aggregated congestion-feasibility proof and show that each radial-feeder grid constraint, thermal line-capacity limits and bus voltage-magnitude bounds alike, can be reconstructed homomorphically from per-bus Pedersen commitments, reducing all feasibility checks to a single aggregated range proof whose size grows logarithmically with the number of constraints.
\item We propose congestion-responsibility fairness, a new metric to measure fairness grounded in grid physics, which measures whether curtailment reflects each party's physical contribution to congested lines, rather than consumption or equal division. It captures disparities missed by conventional measures such as the Gini coefficient.
\item We reproduce three published privacy-preserving and grid-feasible trading designs and compare them with \name on common market rounds, real distribution feeders, and a $240$-node feeder. We further evaluate \name on a live Hyperledger Fabric network using Hyperledger Caliper.
\end{itemize}

The rest of this paper is organized as follows: Section~\ref{sec:background} provides background on P2P energy-trading, blockchain and zk proofs. 
% Section~\ref{sec:related} reviews related work, 
Section~\ref{sec:technical_details} presents the system overview and technical details of \name, Section~\ref{sec:eval} evaluates its effectiveness, and~\ref{sec:conclusion} discuss limitations and conclude the paper.
%%%%%%%%%%%%%%%%%%%%%%%%%%%%%%%%%%%%%%%%%%%%

%%%%%%%%%%%%%%%%% Background %%%%%%%%%%%%%
\vspace{-4pt}

\section{Background}
\label{sec:background}

\vspace{-3pt}

Understanding energy trading with \name requires familiarity with blockchain, ZK proofs, and cryptographic commitments. This section provides the necessary background.

\vspace{-2pt}
\subsection{P2P Energy Markets and Network-Constrained Clearing}
\label{subsec:p2p}
\vspace{-2pt}
A P2P local energy market lets prosumers trade surplus generation directly over short, repeated clearing intervals, and is increasingly settled on a distributed ledger for tamper-evident, auditable records of bids, matches, and settlements~\cite{sousa2019peer}. An energy market is constrained by physics: traded power flows over shared distribution feeders subject to hard thermal and voltage limits, therefore a market outcome is admissible only if it is also \emph{network-feasible}, which state-of-the-art clearing enforces by embedding a linearized power-flow model into the matching logic~\cite{guerrero2018decentralized}. When a feeder segment is congested, its remaining capacity becomes a scarce resource shared by competing trades. Thus, the order in which trades are considered can determine which trades receive the remaining capacity and which are curtailed (more in~\ref{subsec:fabric}).

\vspace{-4pt}
\subsection{Consortium Blockchains and the Hyperledger Fabric}
\label{subsec:fabric}
\vspace{-2pt}

\begin{figure}[t]
    \centering
    \includegraphics[width=.75\columnwidth]{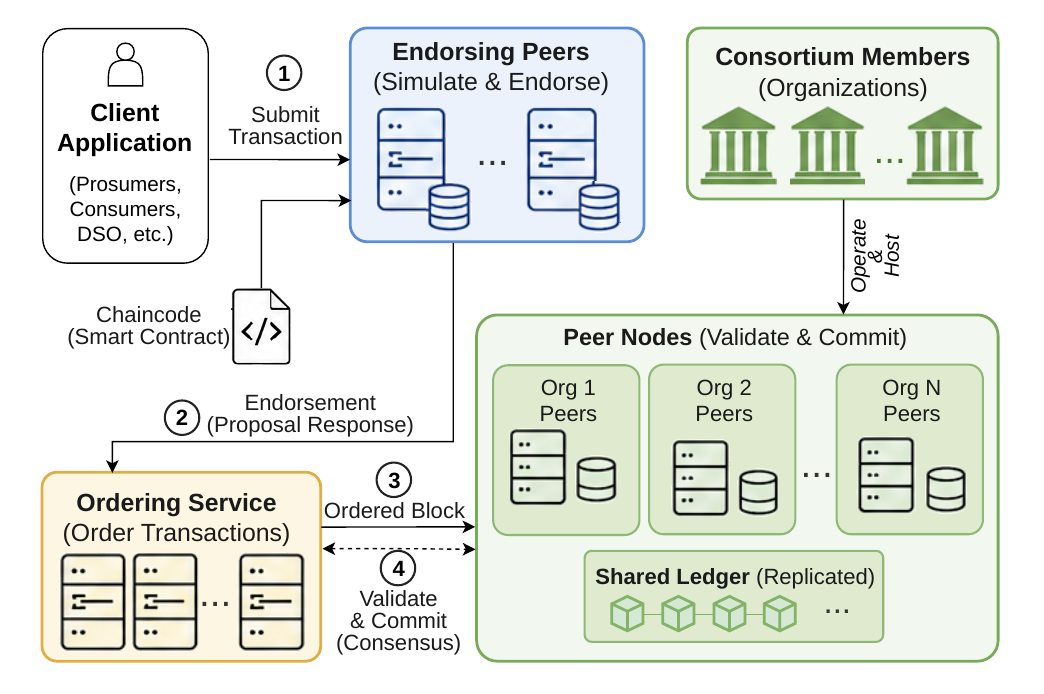}
    \vspace{-8pt}
    \caption{Consortium Permissioned Blockchain.}
    \vspace{-15pt}
    \label{fig:blockchain}
\end{figure}

Energy markets are naturally permissioned: the DSO, retailers, and registered prosumers are known and vetted. Hyperledger Fabric, a widely used consortium blockchain (Figure~\ref{fig:blockchain}) framework, follows an execute-order-validate model in which a distinct \emph{ordering service} determines the total transaction sequence that downstream validation preserves~\cite{androulaki2018hyperledger}. Consensus among orderer nodes guarantees agreement on \emph{a} sequence. However, it does not determine \emph{which} one. The entity best positioned to run that service is the DSO. Combined with the physical scarcity mentioned in Section~\ref{subsec:p2p}, this ordering privilege becomes an allocation privilege over congested feeder.

%%%%%%%%%%%%%%%%%%%%%%%%%%%%%%%%
%%%%%%%%%%%%%%%%%%%%%%%%%%%%%%%%

\subsection{Commitments and Zero-Knowledge Proofs}
\label{subsec:zkp}
\vspace{-2pt}
To keep prosumers' bids private and to prevent participants from reacting to one another's offers, these markets increasingly adopt cryptographic hiding. A \emph{commitment scheme} lets a party publish a commitment to a value (e.g., a bid) that reveals nothing about the value yet cannot later be opened to a different value. Pedersen commitments are a standard instantiation~\cite{pedersen1991non}: given a group $\mathbb{G}$ of prime order $q$ with generators $g,h$ whose discrete logarithm $\log_g h$ is unknown, a bid $m\in\mathbb{Z}_q$ is committed with fresh randomness $r\in\mathbb{Z}_q$ as

\vspace{-12pt}

\begin{equation}
\mathsf{Com}(m,r) = g^{m} h^{r}
\label{eq:pedersen}
\end{equation}

\vspace{-6pt}

This commitment is perfectly hiding and computationally binding under the discrete-logarithm (DL) assumption, and it is additively homomorphic,

\vspace{-18pt}

\begin{equation}
\mathsf{Com}(m_1,r_1)\cdot\mathsf{Com}(m_2,r_2)
= \mathsf{Com}(m_1+m_2,\; r_1+r_2)
\label{eq:homomorphism}
\end{equation}

\vspace{-4pt}

which allows committed quantities to be aggregated without opening them. A sealed-bid auction is then run as a two-phase \emph{commit-reveal} protocol: participants first publish commitments Eq.~\eqref{eq:pedersen} to their bids and only afterward reveal the underlying values, therefore no participant can tailor a bid to another's.
\vspace{-4pt}
%%%%%%%%%%%%%%%%%%%%%%%%%%%%%%%%%%%%%%%%%%%%

% %%%%%%%%%%%%%%%%% Related Work %%%%%%%%%%%%%
% \input{Sections/related_work}
% %%%%%%%%%%%%%%%%%%%%%%%%%%%%%%%%%%%%%%%%%%%%

%%%%%%%%%%%%%%%%% System Overview %%%%%%%%%%%%%
% \input{Sections/overview}
%%%%%%%%%%%%%%%%%%%%%%%%%%%%%%%%%%%%%%%%%%%%

%%%%%%%%%%%%%%%%% Technical Details %%%%%%%%%%%%%
\section{Technical Details}\label{sec:technical_details}
\vspace{-4pt}

This section provides an overview of \name, the threat model which \name defends against, and discusses core technical elements of the proposed P2P energy trading system. Table~\ref{tab:notation} summarizes the notation used throughout this section.

\subsection{System Overview}
\label{sec:overview}
\vspace{-4pt}

%%%%%%%%%%%%%%%%%%%%
\begin{table}[!t]
\centering
\caption{Summary of notation.}
\vspace{-5pt}
\setlength{\tabcolsep}{3pt}
\renewcommand{\arraystretch}{1.05}
\scriptsize
\begin{tabular}{|c|>{\raggedright\arraybackslash}p{3.15cm}|c|>{\raggedright\arraybackslash}p{2.7cm}|}
\hline
\textbf{Symbol} & \textbf{Meaning} & \textbf{Symbol} & \textbf{Meaning} \\
\hline
$n,j$ & Number of buses, bus index & $c_b,c_s$ & Limit prices \\
\hline
$p_j,\bm p$ & Net injection, injection vector & $g$ & Commitment generator \\
\hline
$K,k$ & Number of constraints, index & $C_j$ & Commitment to $p_j$ \\
\hline
$F,f_k$ & Constraint matrix, $k$-th row & $\omega_j$ & Blinding factor of $C_j$ \\
\hline
$\underline b_k,\overline b_k$ & Lower, upper bound of constraint $k$ & $\Omega_k$ & Aggregated blind of constraint $k$ \\
\hline
 $f_{kj}$ & Coefficient of bus $j$ in row $k$ & $d_k^{\text{lo}},d_k^{\text{hi}}$ & Floor, ceiling margins \\
\hline
$m,\ell$ & Number of lines, line index & $L_k,U_k$ & Margin commitments \\
\hline
% $S,S_{\ell j}$ & Flow sensitivity (downstream indicator) & $\rho_i,\gamma_i$ & Responsibility, curtailment shares \\
$S,S_{\ell j}$ & Flow sensitivity & $\rho_i$ & Responsibility shares \\
\hline
$B$ & Voltage sensitivity & $\gamma_i$ & Curtailment shares \\
% $S,S_{\ell j}$ & Flow sensitivity & \multirow{2}{*}{$\rho_i,\gamma_i$} & \multirow{2}{2.7cm}{Responsibility, curtailment shares} \\
% \cline{1-2}
% $B$ & Voltage sensitivity & & \\
\hline
$\limit_\ell$ & Rating of line $\ell$ & $\beta$ & Range-proof bit-width \\
\hline
$\mathcal B,\mathcal S$ & Buy, sell bid sets & $\mathcal L^{*}$ & Binding lines \\
\hline
$b,s$ & Buy, sell bid index & $i,\mathrm{bus}(i)$ & Party index, its bus \\
\hline
$W^\star,W_{\mathrm s}$ & Optimal, settled welfare & $d_i$ & Quantity of party $i$ \\
\hline
$\pi$ & Aggregated range proof & $r_i$ & Congestion responsibility \\
\hline
$x_{bs},\bm x$ & Traded quantity, trade vector & $d_b,d_s$ & Bid quantities \\
\hline
$A$ & Trade-to-bus incidence matrix & $\mathrm{curt}_i$ & Curtailment of party $i$ \\
\hline
\end{tabular}
\vspace{-15pt}
\label{tab:notation}
\end{table}
%%%%%%%%%%%%%%%%%%%%

\begin{figure}[t]
\centering
% \vspace{-8pt}
\includegraphics[width=\columnwidth]{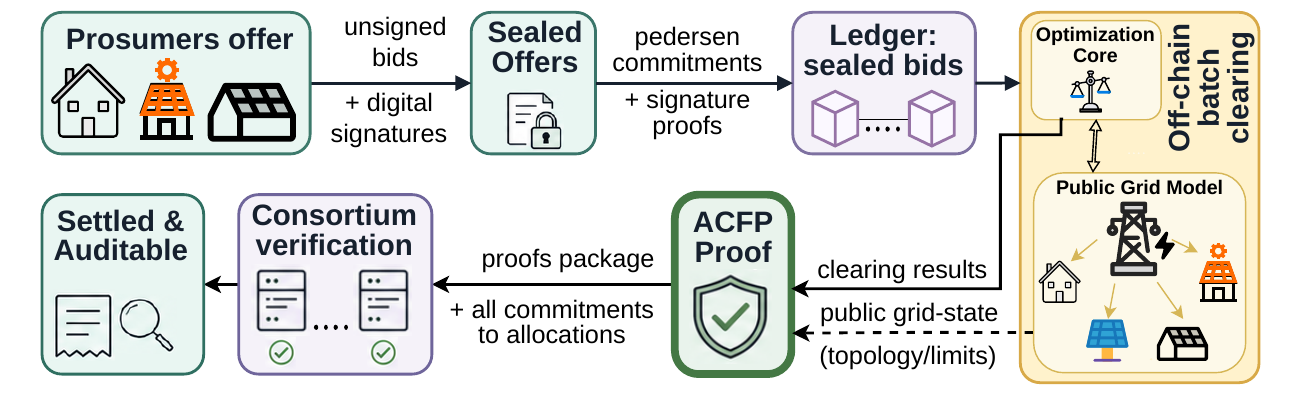}
\vspace{-20pt}
\caption{\name System Overview}
\vspace{-8pt}
\label{fig:overview}
\end{figure}

This section follows one market round end to end (Figure~\ref{fig:overview}). Prosumers submit buy/sell offers, each naming a price, quantity, and bus; every offer is \emph{sealed} as a commitment and recorded on the ledger, whose bid-set hash fixes exactly which offers belong to the round. Once the round closes, the operator clears off-chain with an \emph{order-independent} batch linear program (LP) that maximises welfare subject to the public grid model, so reordering trades gains nothing. It must then convince every peer that the allocation is grid-safe without revealing it: this is ACFP, which rebuilds each grid-constraint commitment from the per-bus commitments and proves feasibility, over the feeder's public thermal line-capacity limits and bus voltage-magnitude bounds in a single aggregated range proof. Each peer verifies independently and commits only if its own check passes, so no party can assert feasibility unilaterally. Settlement remains auditable: any participant may challenge the round and force a reveal-and-recompute, catching a feasible-yet-unfair clearing.

%%%%%%%%%%%%%%%%%%%%%%%%%%%%%%%%%%%%%%%%%%%%%%%%%%%%%%%%%%%
\vspace{-4pt}
\subsection{Grid Model and Threat Model}
\vspace{-4pt}

% A radial distribution feeder serves prosumers who trade energy peer-to-peer over a consortium blockchain. Each round, prosumers submit buy or sell \emph{bids} that a market \emph{operator}, typically the DSO, clears. A clearing induces a net injection
% $p_j$ at every bus $j$, written as a vector $\bm p\in\mathbb Z^{n}$ over $n$ buses,
% where $p_j>0$ means bus $j$ injects power and $p_j<0$ means it draws power.

A radial distribution feeder serves prosumers who trade energy peer-to-peer over a consortium blockchain; each round, an operator, typically the DSO, clears buy or sell bids into a net injection $p_j$ (real-power) at every bus $j$, with $p_j>0$ injecting and $p_j<0$ drawing power. Collecting these into $\bm p\in\mathbb Z^{n}$ over $n$ buses gives the allocation whose feasibility ACFP must certify. \name treats grid feasibility \emph{abstractly}, as an arbitrary set of $K$ \emph{public linear constraints} on the injection vector. Collecting their coefficients into a public integer matrix $F\in\mathbb Z^{K\times n}$ with per-constraint integer bounds $\underline b_k,\overline b_k$, a clearing is \emph{grid-feasible} exactly when:
\vspace{-6pt}
%%%%%%%%%%%%%%%%%%%%%%%%%%%%%%%%%%%%%%%%%%%
\begin{equation}\label{eq:feas}
\underline b_k\;\le\;(F\bm p)_k\;\le\;\overline b_k,\qquad k=1,\dots,K.
\vspace{-4pt}
\end{equation}
%%%%%%%%%%%%%%%%%%%%%%%%%%%%%%%%%%%%%%%%%%%
Each row $f_k$ of $F$ is a public linear functional of the injections; ACFP (Section~\ref{sec:acfp}) certifies all $K$ of them in a single proof. Any grid limit expressible as a linear (or linearized) function of the nodal injections is admissible, and a new limit is added simply by appending a row to $F$, with no change to the protocol. For our evaluation we instantiate $F$ with thermal and voltage limits, the two most common distribution-feeder limits, both from the linearized DistFlow model~\cite{wu1989network}. On a radial tree the real-power flow on a line equals the sum of the net injections downstream of it, so the flow sensitivity is a $0/1$ indicator over the $m$ lines,

\vspace{-16pt}
%%%%%%%%%%%%%%%%%%%%%%%%%%%%%%%%%%%%%%%%%%%
\begin{equation}\label{eq:sensitivity}
S_{\ell j}=\mathbf 1[\,\text{\footnotesize bus }j\text{\footnotesize\ downstream of line }\ell\,],\quad S\in\{0,1\}^{m\times n},
\vspace{-4pt}
\end{equation}
%%%%%%%%%%%%%%%%%%%%%%%%%%%%%%%%%%%%%%%%%%%
and the rating $|(S\bm p)_\ell|\le\limit_\ell$ is the symmetric interval $[-\limit_\ell,\,\limit_\ell]$.

Bus-voltage limits fit the same mould: under LinDistFlow the squared bus voltage is likewise a public linear function of the injections, $(B\bm p)_b$, through a public voltage-sensitivity $B$ (radial feeder, public line impedances, and a fixed public power factor folding in the reactive drop), so a voltage band is a two-sided interval on $(B\bm p)_b$. Stacking the thermal and voltage rows gives $F=[S;B]$, so Eq.~\eqref{eq:feas} carries both limits at once.

% The second block is the bus-voltage limit. LinDistFlow renders the squared bus voltage affine in the injections through a public sensitivity

% \begin{equation}\label{eq:voltage}
% B = 2\,S^{\!\top}\!\operatorname{diag}(\bm r + \bm x\tan\phi)\,S,
% \end{equation}

% %%%%%%%%%%%%%%%%%%%%%%%%%%%%%%%%%%%%%%%%%%%
% % \begin{equation}\label{eq:voltage}
% % B=2\,S^{\!\top}\!\operatorname{diag}(\bm r+\bm x\tan\phi)\,S,
% % \end{equation}
% %%%%%%%%%%%%%%%%%%%%%%%%%%%%%%%%%%%%%%%%%%%

% built from the topology $S$ and public line impedances $(\bm r,\bm x)$, with a fixed public power factor $\cos\phi$ folding the reactive drop into an effective resistance so that $\bm p$ stays the only private vector. The voltage band is then a two-sided interval on $(B\bm p)_b$, the form of Eq.\eqref{eq:feas}. Stacking the thermal and voltage rows gives $F=[S;B]$, so Eq.\eqref{eq:feas} carries both limits at once.

\kafi{We model a compromised operator, an adversary that has captured the ordering role (Section~\ref{subsec:fabric}), whether through a compromised host, an insider, or a hijacked ordering service, and so controls both transaction ordering and market clearing. It may (i)~reorder trades for profit, (ii)~settle an allocation that violates a grid constraint to favor an ally, (iii)~tamper with the recorded allocation, (iv)~fabricate congestion to curtail a rival, or (v)~settle a feasible yet deliberately sub-optimal clearing. We assume consortium peers verify every round independently, an honest peer majority, hardness of the DL problem, and no trusted setup.}

%%%%%%%%%%%%%%%%%%%%%%%%
%%%%%%%%%%%%%%%%%%%%%%%%

%%%%%%%%%%%%%%%%%%%%%%%%%%%%%%%
\vspace{-4pt}
\subsection{Order-Independent, Welfare-Optimal Market Clearing}\label{sec:clearing}
\vspace{-2pt}

Let $\mathcal B$ and $\mathcal S$ denote the buy and sell bids, where bid $b$ carries a limit price $c_b$ and a quantity $d_b$. Instead of matching bids one at a time, we clear the whole round with a single LP. The decision variables $x_{bs}\ge 0$ range only over \emph{price-compatible} pairs, those with $c_b\ge c_s$; a pair that could never profitably trade never enters the program. The map from trades to nodal injections is a fixed \emph{incidence matrix} $A\in\{-1,0,+1\}^{n\times|\bm x|}$, whose column for trade $(b,s)$ places $+1$ at the seller's bus and $-1$ at the buyer's. This gives $\bm p=A\bm x$, so the grid constraints Eq. \eqref{eq:feas} act on $FA\bm x$. The clearing solves
\vspace{-8pt}
%%%%%%%%%%%%%%%%%%%%%%%%%%%%%%%%%%%%%%%%%%%
\begin{subequations}\label{eq:lp}
\begin{align}
\max_{\bm x\ge\bm 0}\quad & \textstyle\sum_{(b,s)} x_{bs}\,(c_b-c_s) \label{eq:lp-obj}\\
\text{s.t.}\quad & \textstyle\sum_{s} x_{bs}\le d_b \quad\forall b\in\mathcal B,
                   \label{eq:lp-demand}\\
                 & \textstyle\sum_{b} x_{bs}\le d_s \quad\forall s\in\mathcal S,
                   \label{eq:lp-supply}\\
                 & \underline{\bm b}\;\preceq\; F\,A\,\bm x\;\preceq\;\overline{\bm b}.
                   \label{eq:lp-grid}
\end{align}
\vspace{-0pt}
\end{subequations}
%%%%%%%%%%%%%%%%%%%%%%%%%%%%%%%%%%%%%%%%%%%
% \vspace{-5pt}
The objective Eq. \eqref{eq:lp-obj} is social welfare; constraints Eq. \eqref{eq:lp-demand}-\eqref{eq:lp-supply} cap each buyer and seller at what it asked for or offered, and Eq. \eqref{eq:lp-grid} carries the entire constraint set Eq. \eqref{eq:feas}, making every grid limit part of the market clearing.

Every input to this LP, the surpluses $c_b-c_s$, the quantities $d_b,d_s$, the bounds $\underline{\bm b},\overline{\bm b}$, and $F,A$, depends only on the bid \emph{set} and public grid data, never on the order bids arrived in, so the order has no bearing on the clearing and single-round ordering MEV is exactly zero. The LP also returns the welfare-maximizing allocation, so no sequential matching of the same bids can do better: reordering gains the operator nothing.

\vspace{-2pt}
\subsection{ACFP: Feasibility over a Hidden Allocation}\label{sec:acfp}
\vspace{-2pt}
The operator must convince every peer that the settled $\bm p$ satisfies Eq. \eqref{eq:feas}, without revealing $\bm p$ or asking anyone to trust its word. ACFP achieves this with the fact that holds for \emph{every} constraint in $F$, whatever it encodes: each $(F\bm p)_k$ is linear in $\bm p$, and Pedersen commitments are additively homomorphic Eq. \eqref{eq:homomorphism}, so a commitment to every constraint's value can be built directly from commitments to the bus injections.

The operator publishes one commitment per bus, $C_j=\Com(p_j,\omega_j)$ Eq. \eqref{eq:pedersen}, together with the settled trades. Since every coefficient $f_{kj}$ is public, raising each $C_j$ to $f_{kj}$ and multiplying gives, by the homomorphism, a commitment to constraint $k$'s value $(F\bm p)_k$:

\vspace{-6pt}
\begingroup
\footnotesize
%%%%%%%%%%%%%%%%%%%%%%%%%%%%%%%%%%%%%%%%%%%
\begin{equation}\label{eq:cflow}
\prod_{j=1}^{n} C_j^{\,f_{kj}} = \Com\big((F\bm p)_k,\;\Omega_k\big), \qquad \Omega_k=\textstyle\sum_j f_{kj}\,\omega_j,
\end{equation}
%%%%%%%%%%%%%%%%%%%%%%%%%%%%%%%%%%%%%%%%%%%
\endgroup

therefore, each verifier computes Eq. \eqref{eq:cflow} itself from public data, rather than trusting a value from the operator; the coefficients $f_{kj}$ may be arbitrary integers. To show $(F\bm p)_k$ lands inside its bounds without opening the commitment, we ask about two ordinary, non-negative numbers: \emph{how far above the floor} it sits, and \emph{how far below the ceiling} it sits,
\vspace{-4pt}
\[
d_k^{\text{lo}} := (F\bm p)_k-\underline b_k,\qquad\qquad
d_k^{\text{hi}} := \overline b_k-(F\bm p)_k.
\vspace{-4pt}
\]
The constraint $\underline b_k\le(F\bm p)_k\le\overline b_k$ holds \emph{exactly when both margins are non-negative}. And since the two margins always add up to the same public number, $d_k^{\text{lo}}+d_k^{\text{hi}}=\overline b_k-\underline b_k$.

Each margin is still linear in $\bm p$, it is just $(F\bm p)_k$ shifted by a known constant, so each has its own commitment, built the same homomorphic way as Eq. \eqref{eq:cflow}, only with an extra public factor for the shift:

\vspace{-12pt}
%%%%%%%%%%%%%%%%%%%%%%%%%%%%%%%%%%%%%%%%%%%
\begingroup
\footnotesize
\begin{subequations}\label{eq:vshift}
\begin{align}
{L_k := g^{-\underline b_k}\prod_{j} C_j^{\,f_{kj}} = \Com(d_k^{\text{lo}},\Omega_k),\qquad} \\
U_k := g^{\,\overline b_k}\prod_{j} C_j^{-f_{kj}} = \Com(d_k^{\text{hi}},-\Omega_k).
\end{align}
\end{subequations}
\endgroup
%%%%%%%%%%%%%%%%%%%%%%%%%%%%%%%%%%%%%%%%%%%

$L_k$ commits to the ``above the floor'' margin and $U_k$ to the ``below the ceiling'' margin; proving \emph{both} of them non-negative is the pair of checks that certifies $\underline b_k\le(F\bm p)_k\le\overline b_k$.

The operator produces one aggregated Bulletproofs range proof~\cite{bunz2018bulletproofs} $\pi$, showing all $2K$ margins across all $K$ constraints are non-negative at once, made non-interactive via Fiat-Shamir with a shared bit-width $\beta$, and posts $(\{C_j\},\pi)$. Each peer independently recomputes $\{L_k,U_k\}$ from the published $\{C_j\}$ and the public $F$, verifies $\pi$ against them, and endorses the round only if its own check passes. If any constraint is actually violated, one of its two margins is negative and $\pi$ cannot exist for it: an infeasible clearing has no proof.

\begin{figure}[!t]
\centering
\includegraphics[width=\columnwidth]{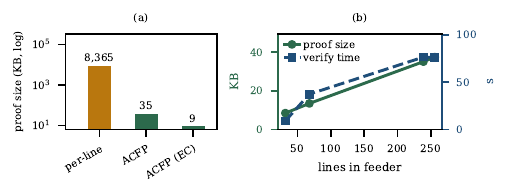}
\vspace{-30pt}
\caption{(a) Proof aggregation impact on proof size and (b) Scaling impact on proof size and verification time with respect to number of lines in feeder.}
\vspace{-15pt}
\label{fig:proofsize}
\end{figure}
%%%%%%%%%%%%%%%%%%%%%%%%%%%%%%

\vspace{-4pt}
\subsection{Enforcing a Fair Clearing}\label{sec:fair}

\vspace{-4pt}

Feasibility is not fairness: an operator can respect every line limit while curtailing a rival through a feasible yet sub-optimal clearing, and ACFP will still prove it safe. We close this gap with \emph{reveal-and-recompute}: any participant may challenge a round, forcing the operator to reveal the bids and commitment blinds, so the audit recomputes $\Com(p_j,\omega_j)$ against each $C_j$, checks every settled trade references a legitimate opposite-side bid, and re-solves Eq. \eqref{eq:lp} on the revealed bids to recover the optimum $W^\star$. Binding rules out passing this audit under a sub-optimal clearing, so $W_{\mathrm s}<W^\star$ yields a fraud proof of the withheld welfare $W^\star-W_{\mathrm s}$, checked, like an optimistic rollup, only when challenged.

% Feasibility is not fairness: an operator can respect every line limit while still curtailing a rival through a feasible yet sub-optimal clearing, and ACFP will prove it safe. We close this gap with \emph{reveal-and-recompute}: any participant may challenge a round, forcing the operator to reveal the bids and commitment blinds. The audit recomputes $\Com(p_j,\omega_j)$ against each $C_j$, checks that every settled trade references a legitimate opposite-side bid, and re-solves Eq. \eqref{eq:lp} on the revealed bids to recover the optimum $W^\star$. Binding rules out passing this audit under a sub-optimal clearing, so a settled welfare $W_{\mathrm s}<W^\star$ yields a fraud proof of the withheld welfare $W^\star-W_{\mathrm s}$; like an optimistic rollup, the check runs only when challenged.

\textit{Congestion-Responsibility Fairness (CRF).} Curtailment should track how much each party loads the binding lines $\mathcal L^{*}$, not what it consumes. Party $i$'s responsibility is $r_i=\sum_{\ell\in\mathcal L^{*}}|S_{\ell,\mathrm{bus}(i)}|\,d_i$; comparing its share $\rho_i=r_i/\sum_k r_k$ against its curtailment share $\gamma_i=\mathrm{curt}_i/\sum_k\mathrm{curt}_k$ gives
\vspace{-3pt}
%%%%%%%%%%%%%%%%%%%%%%%%%%%%%%%%%%%%%%%%%%%
\begin{equation}\label{eq:crf}
\mathrm{CRF}=1-\tfrac12\sum_i\big|\gamma_i-\rho_i\big|\;\in[0,1],
\vspace{-6pt}
\end{equation}
%%%%%%%%%%%%%%%%%%%%%%%%%%%%%%%%%%%%%%%%%%%

equal to $1$ when curtailment tracks responsibility and $0$ under maximal discrimination. Unlike the Gini coefficient, CRF asks whether the \emph{right} parties bear it.
\vspace{-12pt}
%%%%%%%%%%%%%%%%%%%%%%%%%%%%%%%%
\begin{figure}[hbt!]
    \centering
    \includegraphics[width=0.9\columnwidth, trim={0 0 0 10}, clip]{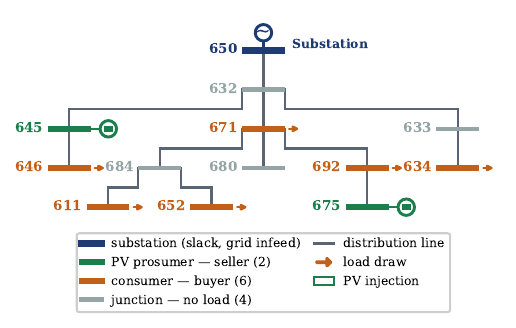}
    \vspace{-15pt}
    \caption{IEEE 13-BUS Topology}
    \vspace{-15pt}
    \label{fig:topology}
\end{figure}
%%%%%%%%%%%%%%%%%%%%%%%%%%%%%%%%
%%%%%%%%%%%%%%%%%%%%%%%%%%%%%%%%%%%%%%%%%%%%

%%%%%%%%%%%%%%%%% Evaluation %%%%%%%%%%%%%
\section{Evaluation}\label{sec:eval}
\vspace{-5pt}

We evaluate \name along four axes: ACFP's proof cost and how it scales with feedersize, the fairness of curtailment, a head-to-head comparison with existing designs, and cost on a live permissioned ledger. Results are for a single clearing round, the most-congested (peak-load) hour of the Midwest 240-node feeder's year of real hourly loads, with 193 sealed bids clearing into 113 trades; the smallest of our feeders, the IEEE 13-bus system, is shown in Figure~\ref{fig:topology} with each bus's market role annotated. Commitments use a 1024-bit safe-prime group, so the reported cryptographic times are conservative relative to a 256-bit elliptic-curve instantiation (Figure~\ref{fig:proofsize}). Based on our experiments, on a plaintext continuous double auction, \kafi{a compromised} operator can swing total welfare by 17\% (Figure~\ref{fig:welfalre_distribution}) simply by reordering an honest bid set. \name's batch clearing (Section~\ref{sec:clearing}) drives this to exactly zero, however, leaves room for infeasible distribution.

% \vspace{-15pt}

% by construction, since the outcome is a function of the bid \emph{set} alone. Order-independence still leaves \kafi{a compromised} operator free to announce an allocation the network cannot carry, which is what the rest of this section addresses.

%%%%%%%%%%%%%%%%%%%%%%%%%%
\begin{figure}[!t]
\centering
\includegraphics[width=\columnwidth]{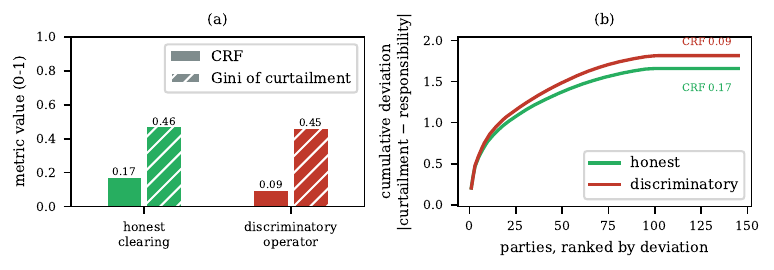}
\vspace{-25pt}
\caption{(a) Fairness comparison between CRF and Gini Coefficient and (b) Impact of scaling deviation on the gap between CRF and Gini Coefficient.}
\vspace{-10pt}
\label{fig:fairness}
\end{figure}
%%%%%%%%%%%%%%%%%%%%%%%%%%

\vspace{-4pt}

\subsection{Proof size and scalability}\label{sec:eval-proof}

\vspace{-3pt}
% Aggregating the feeder's 239 line constraints into a single range argument cuts proof size from 8{,}365~KB to \textbf{35~KB}, a $\approx$238$\times$ reduction (Figure~\ref{fig:proofsize}a) that only widens as more constraints are stacked, since the aggregated argument is logarithmic in the constraint count; a 256-bit elliptic curve would shrink this further, to roughly 8.8~KB. Across feeders (Figure~\ref{fig:proofsize}b), the range argument grows logarithmically with the number of lines while the per-bus commitments grow linearly, so overall proof size scales sub-linearly: an 8$\times$ larger constraint set (IEEE 123-node vs.\ 33-bus) costs only 4.4$\times$ more bytes. The Midwest feeder's proof is smaller than the 123-node feeder's despite having more nodes, since per-phase LinDistFlow~\cite{wu1989network} expands the 123-node feeder into more composite (node, phase) buses, and bus count, not line count, drives the linear term. Because ACFP reconstructs flows homomorphically, the constraint matrices never become arithmetic-circuit constraints, so ACFP needs no circuit-specific trusted setup where a generic SNARK would.
Aggregating the feeder's 239 line constraints into a single range argument cuts proof size from 8{,}365~KB to \textbf{35~KB}, a $\approx$238$\times$ reduction (Figure~\ref{fig:proofsize}a), and because the aggregated argument is logarithmic in the constraint count, the reduction only widens as more constraints are stacked. A 256-bit elliptic curve would shrink this further, to roughly 8.8~KB. Across feeders (Figure~\ref{fig:proofsize}b), the range argument grows logarithmically with the number of lines while the per-bus commitments grow linearly, so overall proof size scales sub-linearly: an 8$\times$ larger constraint set (IEEE 123-node vs.\ 33-bus) costs only 4.4$\times$ more bytes. The Midwest feeder's proof is smaller than the 123-node feeder's despite having more nodes, since per-phase LinDistFlow~\cite{wu1989network} expands the 123-node feeder into more composite (node, phase) buses. It is bus count that drives the linear term. Moreover, since ACFP reconstructs flows homomorphically, the constraint matrices never become arithmetic-circuit constraints, therefore ACFP needs no circuit-specific trusted setup where a generic SNARK would.

\vspace{-4pt}

% %%%%%%%%%%%%%%%%%%%%%%%%%%%%%%%%
% \begin{figure}[t]
%     \centering
%     \includegraphics[width=.95\columnwidth]{Figures/05_scaling.pdf}
%     \vspace{-8pt}
%     \caption{Proof size and per-peer verification time against feeder size}
%     \vspace{-15pt}
%     \label{fig:scaling}
% \end{figure}
% %%%%%%%%%%%%%%%%%%%%%%%%%%%%%%%%

% \vspace{-2pt}

\subsection{Fairness of curtailment}\label{sec:eval-fairness}

\vspace{-4pt}

% Under a clearing where \kafi{a compromised} operator favors a colluding ally, CRF Eq. \eqref{eq:crf} drops from \textbf{0.17} to \textbf{0.09}, a 46\% relative fall (Figure~\ref{fig:fairness}), while the curtailment Gini coefficient barely moves, from 0.465 to 0.455. Roughly the same amount of curtailment concentrates on roughly the same number of parties in both clearings, therefore Gini cannot see \emph{who} bears it, while CRF can, since it compares curtailment against physical responsibility rather than concentration. The honest clearing's own CRF (0.17) sits far below the ideal of 1.0 since the LP maximizes welfare, not proportionality. It curtails whoever is cheapest, not necessarily whoever loads the line most. CRF is therefore a comparative detector rather than an absolute certificate, one that moves sharply and predictably under favoritism, making a persistent cross-round drop actionable evidence for a regulator. Figure~\ref{fig:discrimination} confirms this directly. Shifting a growing share of curtailment from the most-responsible to the least-responsible party, holding total curtailment fixed, drives CRF down linearly on every feeder while the curtailment Gini stays flat or moves the wrong way. How steeply CRF responds depends on how concentrated curtailment already is, where one party bears a large share (the small IEEE 13-bus and Midwest feeders), reallocating it moves CRF sharply. where curtailment is already spread thin (the 123-node feeder), the drop is hardly noticeable.

\begin{figure}[t]
    \centering
    \hspace{-0.08\columnwidth}
    \subfloat[]{%
        \includegraphics[width=0.55\columnwidth, trim=0 -14 0 0]{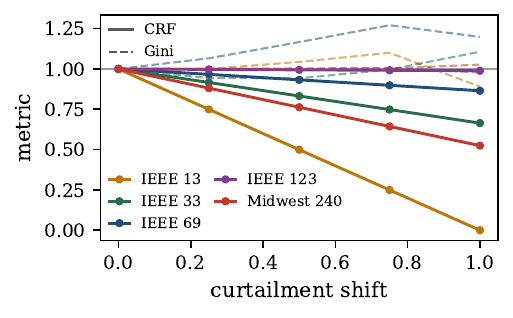}%
        \vspace{-25pt}
        % \caption{CRF vs. Gini across feeders}
        % \vspace{-10pt}
        \label{fig:discrimination}
    }
    \hspace{-0.04\columnwidth}
    % \hfill
    \subfloat[\centering ]{%
        \includegraphics[width=0.50\columnwidth]{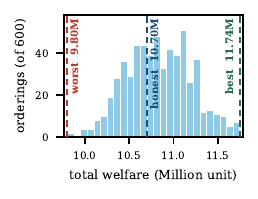}%
        \vspace{-25pt}
        % \caption{\centering Transaction Order's Impact on Total Welfare}
        % \vspace{-10pt}
        \label{fig:welfalre_distribution}
    }
    \vspace{-5pt}
    \caption{(a) CRF vs. Gini across feeders and (b) Transaction Order's Impact on Total Welfare.}
    \vspace{-12pt}
    \label{fig:overall}
\end{figure}

Under a clearing where \kafi{a compromised} operator favors a colluding ally, CRF Eq. \eqref{eq:crf} drops from \textbf{0.17} to \textbf{0.09}, a 46\% relative fall (Figure~\ref{fig:fairness}), while the curtailment Gini coefficient barely moves, from 0.465 to 0.455: Gini misses \emph{who} bears curtailment, since roughly the same share falls on roughly the same parties, while CRF, weighing curtailment against physical responsibility, does not. The honest clearing's own CRF (0.17) sits well below 1.0, since the LP maximizes welfare, not proportionality. Figure~\ref{fig:discrimination} confirms this: shifting curtailment from the most- to least-responsible party drives CRF down linearly on every feeder while Gini stays flat, sharpest where curtailment concentrates (IEEE 13-bus, Midwest), negligible where it is spread thin (123-node).

% \begin{figure}[!t]
% \centering
%     \begin{subfigure}[t]{0.49\columnwidth}
%         \centering
%         \includegraphics[width=1.00\columnwidth]{Figures/10_discrimination_response.pdf}
%         \vspace{-5pt}
%         \caption{CRF vs. Gini across feeders}
%         \vspace{-10pt}
%         \label{fig:discrimination}
%     \end{subfigure}
%     \hfill
%     \begin{subfigure}[t]{0.49\columnwidth}
%         \centering
%         \includegraphics[width=1.00\columnwidth]{Figures/12_welfare_distribution_midwest.pdf}
%         \vspace{-5pt}
%         \caption{CRF vs. Gini across feeders}
%         \vspace{-10pt}
%         \label{fig:dis}
%     \end{subfigure}
% \end{figure}

% \begin{figure}[t]
%     \centering
%     \hspace{-0.08\columnwidth}
%     \begin{minipage}[t]{0.56\columnwidth}
%         \centering
%         \includegraphics[width=\linewidth, trim=0 -18 0 0]{Figures/10_discrimination_response.pdf}
%         \vspace{-25pt}
%         \caption{CRF vs. Gini across feeders}
%         \vspace{-10pt}
%         \label{fig:discrimination}
%     \end{minipage}
%     \hspace{-0.045\columnwidth}
%     \begin{minipage}[t]{0.52\columnwidth}
%         \centering
%         \includegraphics[width=\linewidth]{Figures/12_welfare_distribution_midwest.pdf}
%         \vspace{-25pt}
%         \caption{\centering Transaction Order's Impact on Total Welfare}
%         \vspace{-10pt}
%         \label{fig:welfalre_distribution}
%     \end{minipage}
%     \hspace{-0.04\columnwidth}
% \end{figure}

\vspace{-4pt}

\subsection{Comparison with existing approaches}\label{sec:eval-repro}

\vspace{-2pt}

\begin{figure}[!t]
\centering
\includegraphics[width=\columnwidth]{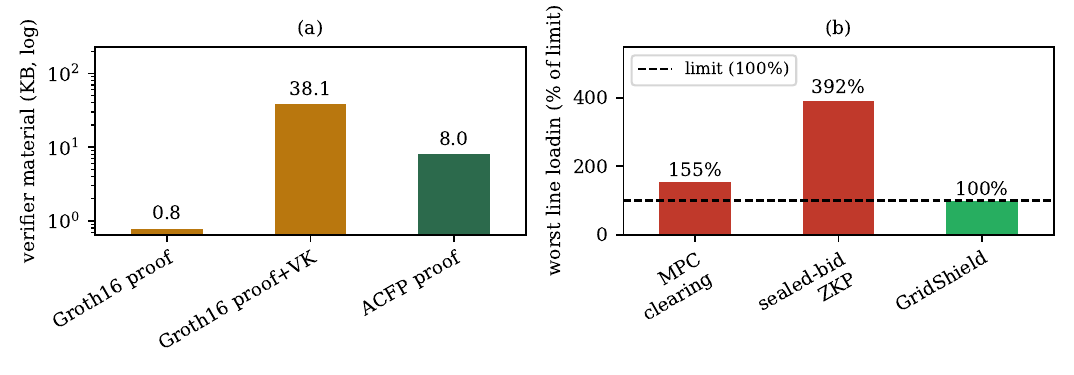}
\vspace{-28pt}
\caption{(a) Total size of information a verifier needs to hold for proof verification and (b) Accepted energy flow allocation.}
\vspace{-15pt}
\label{fig:repro}
\end{figure}

We implemented the load-bearing component of three exisiting approaches: Groth16-based feasibility circuit~\cite{kim2026zero}, MPC secret-shared clearing~\cite{mitrea2023privacy,zhou2023bidirectional}, and Sealed-bid$+$per-match ZKP~\cite{hou2022privacy}, and ran them on our data. Table~\ref{tab:reproduction} and Figure~\ref{fig:repro} summarize the comparison among them.
% , and Table~\ref{tab:quant} places our measurements beside the figures the original papers report.

A Groth16 feasibility circuit~\cite{kim2026zero} emits a smaller proof than ours (0.79~KB vs.\ 8.5~KB), however a verifier must also obtain and trust its 38.1~KB verification key; ACFP has no such key, so its entire verifier-side material is the 8.5~KB proof itself, 4.6$\times$ less than Groth16's combined footprint, with no trusted setup. More important is what each verifier will \emph{accept}: an MPC-based clearing~\cite{mitrea2023privacy,zhou2023bidirectional} that omits the network constraint loads a line to \textbf{155\%} of rating and still verifies; a sealed-bid ZKP design~\cite{hou2022privacy} that checks individual matches without checking checking aggregate volume accepts an allocation we constructed at \textbf{392\%}. \name's proof cannot exist above \textbf{100\%} by construction. We confirmed this feasibility under full AC power flow in OpenDSS.

%%%%%%%%%%%%%%%%%%%%%%%%%%%%%%%%%%%%%%
%%%%%%%%%%%%%%%%%%%%%%%%%%%%%%%%%%%%%%

Table~\ref{tab:reproduction} shows only Groth16 and \name prove feasibility, and only \name does so without a trusted setup or interaction: the MPC approach needs 6{,}592 communication rounds and convinces only the parties who ran it (about 66~s at a realistic round-trip time), whereas \name's 82.2~s prove+verify yields an artefact any third party can check.

\begin{table}[htbp]
\centering
\vspace{-6pt}
\caption{Reproduced prior designs versus \name}
\vspace{-4pt}
\label{tab:reproduction}
\scriptsize
\setlength{\tabcolsep}{4pt}
\renewcommand{\arraystretch}{1.3}
\resizebox{0.98\columnwidth}{!}{%
\begin{tabularx}{\columnwidth}{c|c|c|c|c|c|c}

\textbf{Approach} &
\rotatebox{90}{\textbf{\shortstack[l]{Feasibility\\ proven}}} &
\rotatebox{90}{\textbf{\shortstack[l]{Allocation\\ hidden}}} &
\rotatebox{90}{\textbf{\shortstack[l]{Publicly\\ verifiable}}} &
\rotatebox{90}{\textbf{\shortstack[l]{Trusted\\ setup}}} &
% \rotatebox{90}{\textbf{\shortstack[l]{Interactive}}} &
% \rotatebox{90}{Proof size} &
\rotatebox{90}{\textbf{\shortstack[l]{Prove $+$\\verify (s)}}} &
\rotatebox{90}{\textbf{\shortstack[l]{Infeasible\\ blocked}}} \\
\hline
% \noalign{\vskip 2pt \hrule height 0.4pt width 1.01\columnwidth \vskip 2pt}
Groth16 feasibility circuit \cite{kim2026zero}
  & yes & yes & yes & yes & --- & yes \\
% \hline
MPC secret-shared clearing \cite{mitrea2023privacy,zhou2023bidirectional}
  & no & yes & no & no & 66 & no \\
% \hline
Sealed-bid$+$per-match ZKP \cite{hou2022privacy}
  & no & partial & yes & no & 4.9 & no \\
% \hline
\names
  & yes & yes & yes & no & 82.2 & yes \\
\hline
% \noalign{\vskip 2pt \hrule height 0.4pt width 1.04\columnwidth \vskip 2pt}
\end{tabularx}}%
\vspace{-5pt}
\end{table}

\vspace{-6pt}

\subsection{Cost on a live permissioned ledger}\label{sec:eval-fabric}

\vspace{-4pt}

A protocol that is cryptographically sound is still unusable if the ledger carrying it cannot keep up. We therefore deployed the chaincode on a Hyperledger Fabric v2.5.9 network with two organisations, two endorsing peers and a Raft ordering service, and drove it with Hyperledger Caliper using eight long-lived gateway clients. Table~\ref{tab:fabric} reports the result. Sealed-bid submission tracks the offered rate closely up to roughly 1{,}000~tx/s and then flattens at about 1{,}064~tx/s, with no failed transactions at any rate. Average latency stays below 0.2s until that point and then climbs steeply, reaching 15.03s once the offered load is past saturation. Flat throughput together with rising latency is the signature of a saturated pipeline, therefore 1{,}064~tx/s is the practical capacity of this deployment.

Against that capacity, the protocol's own demands are modest. A market round places 194 transactions on the ledger, which over the one-hour clearing interval is an offered rate of 0.054~tx/s, roughly twenty thousand times below saturation. The two operations specific to \name behave as expected: committing a sealed bid takes 2.10~s end to end, dominated by the block cutting interval rather than by cryptography, while posting the clearing takes 17.75~s, since every endorsing peer independently verifies the full 35~KB aggregated proof before endorsing. That verification cost is the price of \kafi{removing the need to trust} the operator, and it is paid once per round. Reading the clearing back is a local query and costs 0.06~s.

\vspace{-4pt}

\begin{table}[!t]
\centering
\caption{Performance of \names on a Hyperledger Fabric Network}
\vspace{-4pt}
\label{tab:fabric}
\footnotesize
\setlength{\tabcolsep}{4pt}
\renewcommand{\arraystretch}{1.25}
\begin{tabular}{@{}c c c c@{}}
\hline
\multicolumn{4}{@{}l}{\textbf{Sealed-bid submission under increasing load}} \\
\textbf{Offered (tx/s)} & \textbf{Committed (tx/s)} & \textbf{Failed} & \textbf{Latency (s)} \\
\hline
25   & 25.2   & 0 & 0.19 \\
100  & 93.9   & 0 & 0.07 \\
400  & 374.9  & 0 & 0.02 \\
800  & 749.5  & 1 & 0.03 \\
1000 & 936.3  & 0 & 0.09 \\
1200 & 1061.6 & 0 & 4.54 \\
1600 & 1064.1 & 0 & 15.03 \\
\hline
% \multicolumn{4}{@{}l}{\textbf{(b) Per-operation cost}} \\
% \multicolumn{2}{@{}l}{Submit a sealed bid} & \multicolumn{2}{r}{2.10 s} \\
% \multicolumn{2}{@{}l}{Post clearing (peers verify the 35 KB proof)} & \multicolumn{2}{r}{17.75 s} \\
% \multicolumn{2}{@{}l}{Read the clearing back (query)} & \multicolumn{2}{r}{0.06 s} \\
% \hline
% \multicolumn{4}{@{}l}{\textbf{(c) What one market round actually demands}} \\
% \multicolumn{2}{@{}l}{Transactions per round} & \multicolumn{2}{r}{194 tx} \\
% \multicolumn{2}{@{}l}{Required rate over a 15 min interval} & \multicolumn{2}{r}{0.216 tx/s} \\
% \multicolumn{2}{@{}l}{Measured capacity of this deployment} & \multicolumn{2}{r}{1064 tx/s} \\
% \hline
\end{tabular}
\vspace{-20pt}
\end{table}

% \begin{table}[!t]
% \centering
% \vspace{-10pt}
% \caption{Measured cost against values reported by existing works}
% \vspace{-4pt}
% \label{tab:quant}
% \footnotesize
% \setlength{\tabcolsep}{4pt}
% \renewcommand{\arraystretch}{1.25}
% \begin{tabular}{@{}l l r r@{}}
% \hline
% \textbf{Setting} & \textbf{Metric} & \textbf{Ours} & \textbf{Reported} \\
% \hline
% % Range proof & size (B) & 672 & 672 \cite{bunz2018bulletproofs} \\
% % \hline
% IEEE 33-bus & proof (KB) & 8.5 & 0.79 \cite{kim2026zero} \\
%  & verifier material (KB) & 8.5 & 38.1 \cite{kim2026zero} \\
%  & prove (s) & 1.94 & n/a \\
%  & verify (s) & 1.05 & n/a \\
% \hline
% Clearing & MPC rounds & 576 & n/a \\
%  & MPC at 10 ms RTT (s) & 5.8 & n/a \\
%  & end to end (s) & 3.00 & 0.54--1.50 \cite{mitrea2023privacy} \\
% \hline
% \end{tabular}

% \vspace{-20pt}
% \end{table}

%%%%%%%%%%%%%%%%%%%%%%%%%%%%%%%%%%%%%%%%%%%%

%%%%%%%%%%%%%%%%% Conclusion %%%%%%%%%%%%%
\section{Conclusion}
\label{sec:conclusion}
\vspace{-5pt}
\name closes the gap between two lines of work that each conceded what the other provided. It clears a sealed-bid market so that no ordering advantage exists, proves the resulting allocation respects every grid constraint without revealing it, and leaves an unfair clearing open to challenge afterwards. ACFP reconstructs each linear grid constraint homomorphically from per-bus commitments, reducing a whole feeder, over a set of such constraints, to one aggregated range proof with no trusted setup, and CRF gives participants and regulators a way to audit curtailment against physical responsibility rather than consumption alone. Two limits bound these results. ACFP certifies feasibility sets expressible as public \emph{linear} constraints; the thermal and voltage instances we evaluate use LinDistFlow, exact for radial feeders, so nonlinear limits (exact apparent-power ratings, losses, meshed AC) fall outside the present form. The throughput figures reflect a two-organisation network on a single host, not a geographically distributed consortium. \\

\vspace{-10pt}
% Two limits bound these results. ACFP certifies any feasibility set expressible as public \emph{linear} constraints on the injections; the instances we evaluate (thermal and voltage) use the LinDistFlow model, which is exact for radial feeders, so nonlinear limits, exact apparent-power ratings, losses, and meshed alternating-current (AC) feasibility proof, fall outside the proposed system. The throughput figures characterise a two-organisation network on a single host rather than a geographically distributed consortium. Instantiating the commitments on a $256$-bit elliptic curve, and extending the grid model beyond the radial case, are the natural next steps.
%%%%%%%%%%%%%%%%%%%%%%%%%%%%%%%%%%%%%%%%%%%%

%%%%%%%%%%%%%%%%% Acknowledgement %%%%%%%%%%%%%
\section{AI Usage Notes}
\vspace{-4pt}

AI tools were used as assistive tools for polishing the manuscript draft, resolving LaTeX syntax issues, and supporting code generation. All technical content was authored and verified by the authors.
%%%%%%%%%%%%%%%%%%%%%%%%%%%%%%%%%%%%%%%%%%%%

% \bibliographystyle{unsrt}
\bibliographystyle{IEEEtran}
\vspace{-5pt}
\bibliography{References}

\end{document}